\documentclass[11pt, fullpage]{article}

\usepackage{url} 
\usepackage{float}
\usepackage{graphicx}
\usepackage{amsfonts, amssymb, amscd}
\usepackage{amsmath, delarray, theorem}
\usepackage{natbib}
\bibpunct{(}{)}{;}{a}{}{,}
\renewcommand{\cite}{\citep}

\newcommand{\mycap}[1]{\caption{{\it #1}}}
\newcommand{\bitem}{\begin{itemize}}
\newcommand{\eitem}{\end{itemize}}
\newcommand{\benum}{\begin{enumerate}}
\newcommand{\eenum}{\end{enumerate}}
\newcommand{\beqnn}{\begin{eqnarray*}}
\newcommand{\eeqnn}{\end{eqnarray*}}
\newcommand{\beqn}{\begin{eqnarray}}
\newcommand{\eeqn}{\end{eqnarray}}
\newcommand{\normal}[2]{\mbox{N}(#1,#2)}

\newcommand{\E}{\mbox{E}}
\newcommand{\var}{\mbox{Var}}
\newcommand{\cov}{\mbox{Cov}}

\newcommand{\myvec}[1]{\mathbf{#1}}

\newcommand{\bfourmatrix}{\begin{array}({cccc})}
\newcommand{\efourmatrix}{\end{array}}
\newcommand{\bvector}{\begin{array}({c})}
\newcommand{\evector}{\end{array}}

\newcommand{\btab}{\renewcommand{\baselinestretch}{1}\begin{table}[hptb]}
\newcommand{\etab}{\end{table}\renewcommand{\baselinestretch}{1}}
\newcommand{\bfig}{\renewcommand{\baselinestretch}{1}\begin{figure}[hptb]}
\newcommand{\efig}{\end{figure}\renewcommand{\baselinestretch}{1}}

\newcommand{\gvec}[1]{\mbox{\boldmath ${#1}$}}

{\theoremstyle{plain}
  }
{\theoremstyle{plain}
  }
{\theoremstyle{plain}
  \newtheorem{proposition}{Proposition}}
{\theoremstyle{plain}
  }
{\theoremstyle{plain}
  }
{\theoremstyle{plain}
  \newtheorem{theorem}{Theorem}}
{\theoremstyle{plain}
  }
{\theoremstyle{plain}
  }
{\theoremstyle{plain}
  }
{\theoremstyle{plain}
  }
{\theoremstyle{plain}
  }
{\theorembodyfont{\rmfamily}
  }
{\theorembodyfont{\rmfamily}
  }
\usepackage{./mywidefile}
\usepackage{textcomp, subfigure}
\def\T{{\mbox{\rm\tiny T}}}

\begin{document}

\title{Threshold-free Evaluation of Medical Tests for Classification and Prediction: Average Precision versus Area Under the {ROC} Curve }
\author{Wanhua Su$^{\dagger}$ \\ 
Department of Mathematics and Statistics \\
MacEwan University \\
Edmonton, AB, Canada T5J 2P2 \\
\\
Yan Yuan$^{\dagger}$ \\
School of Public Health \\
University of Alberta \\
Edmonton, AB, Canada T6G 1C9\\
\\
Mu Zhu \\
Department of Statistics and Actuarial Science \\
University of Waterloo\\ 
Waterloo, ON, Canada N2L 3G1
}
\date{\today}
\maketitle

\centerline{$^{\dagger}$ These authors contributed equally to this paper.}

\def\blind{0}
\begin{abstract}
When evaluating medical tests or biomarkers for disease classification, the area under the receiver-operating characteristic (ROC) curve is a widely used performance metric that does not require us to commit to a specific decision threshold. For the same type of evaluations, a different metric known as the average precision (AP) is used much more widely in the information retrieval literature. We study both metrics in some depths in order to elucidate their difference and relationship. More specifically, we explain mathematically why the AP may be more appropriate if the earlier part of the ROC curve is of interest. We also address practical matters, deriving an expression for the asymptotic variance of the AP, as well as providing real-world examples concerning the evaluation of protein biomarkers for prostate cancer and the assessment of digital versus film mammography for breast cancer screening.      
\end{abstract}

  
\pagebreak

\section{Introduction}
\label{sec:intro}

There are many different metrics for evaluating diagnostic and screening tests in medical sciences; the book by \citet{rocbk} is an authoritative reference in this field. Commonly used metrics include sensitivity, specificity, positive and negative predictive values, positive and negative diagnostic likelihood ratios, among many others. All the aforementioned metrics require the underlying test to make a binary decision, that is, whether the patient being tested has a certain disease or not. Making such a binary decision usually requires a decision threshold, as the underlying test often gives continuous measurements, such as the serum concentration of a metabolite, or the size and density of a mass seen from a medical image. So, for example, the concentration of serum bilirubin has to exceed a certain level for the test to flag the patient as being likely to have liver dysfunction.  Changing the decision threshold always will affect both the sensitivity and the specificity of the test simultaneously --- in particular, raising one while reducing the other. 

The receiver operating characteristic (ROC) curve traces the tradeoff between sensitivity and specificity as the decision threshold varies. Looking at the entire ROC curve is especially useful when we are not ready to commit to a particular decision threshold. Nonetheless, sometimes we still prefer to have a single, numeric performance metric, for example, when we are evaluating hundreds of potential biomarkers --- comparing hundreds of ROC curves is simply not practical or efficient. The Area Under the ROC Curve (AUC) is a much widely used choice in this context --- tests with larger AUC values are considered more powerful. Thus, researchers may pre-screen hundreds of the potential biomarkers using the AUC, and then compare just a handful of top biomarkers using the entire ROC curve. Again, we refer the readers to the book by \citet{rocbk} and its extensive bibliography for more detailed discussions of these issues.

Despite its popularity, the AUC is not perfect. We have encountered two common criticisms about the AUC:
\bitem
\item[(C1)] Two ``qualitatively different'' tests can quite often turn out to have roughly the same AUC values \citep[e.g.,][]{hand2009measuring, KellyZouROC12}.
\item[(C2)] The ROC curve is not equally relevant everywhere; in practice, the initial part of the curve is much more important \citep{pAUC-mcclish, ROC-review-ZC, baker2001proposed, dodd2003partial, KellyZouROC12}. 
\eitem
A natural response to criticism (C2) is the notion of the ``partial AUC'' --- the area under the ROC curve up to a certain cutoff value \citep[see, e.g.,][]{pAUC-mcclish, pAUC-TZ, jiang1996malignant}. We find the partial AUC somewhat unsatisfying because it depends on a subjective cutoff value, which must be supplied a prior. And, to date, we have known of few effective responses to criticism (C1). 

The information retrieval (IR) community faces a similar problem. Mathematically, the following two questions are equivalent: 
\benum
\item How effective can a retrieval algorithm tell if a document is relevant or not? 
\item How effective can a diagnostic (or screening) test tell if a patient is diseased or not? 
\eenum
However, to answer their question, the IR community tends to rely much more heavily (though not exclusively) on another performance metric known as the average precision (AP) --- see, e.g., \citet{peng} and references therein. Like the AUC, the AP is another single, numeric performance metric that does not require us to commit to a decision threshold priori to the analysis. The relationship between these two threshold-free evaluation metrics seems poorly understood, though, if at all. We studied both metrics out of curiosity. After doing so, we found that, to some extent, the AP actually seems to address both criticisms (C1) and (C2) of the AUC.  
 
We proceed as follows.
In Section~\ref{sec:concept}, we define various quantities of interest using a common set of notations. 
In Section~\ref{sec:theory}, we develop some theoretical insights. More specifically, we introduce two notions, which we call {\em stamina} and {\em momentum}, and show that, for the AUC, stamina and momentum are equally important, whereas for the AP, momentum is more important.  
In Section~\ref{sec:multinomial}, we address some practical issues. In particular, we show how the AP is calculated in practice, and derive its asymptotic variance. 
In Section~\ref{sec:example}, we provide two examples.  
In Section~\ref{sec:implications}, we discuss various practical implications of our result. In particular, we discuss when the AP may be considered a more attractive performance metric than the AUC.

\section{Definitions}
\label{sec:concept}

Suppose there are a total of $n$ subjects, some ($n_1$) of which are diseased (class 1) and the rest ($n_0=n-n_1$) of which are not (class 0). For every subject $i$, a diagnostic (or screening) test produces a score, $x_i$, which we can use to rank (or order) the subjects --- e.g., a high score (large $x_i$) means the subject is more likely to be diseased, and vice versa. In this section, we define various concepts associated with evaluating the effectiveness of such a test. The ultimate objective, of course, is to formally define the AUC and the AP so that they can be studied together. In order to do so, it is convenient to start with the so-called hit curve.

\subsection{The hit curve}

Let $i$ denote the {\em ordered} subject index, that is, $x_1 \geq x_2 \geq ... \geq x_n$. If we threshold the scores at $x_k$, declaring all those with scores $\geq x_k$ to be class-1 and all those with scores $< x_k$ to be class-0, we will have a confusion matrix as displayed in Table~\ref{tab:confusion-raw}, where $d(k)$ is the number of subjects with scores $\geq x_k$, and $\hbar(k)$ is the number of subjects {\em truly belonging to} class-1 out of those {\em declared to be} class-1. Clearly, $\hbar(k)$ is a discrete function, defined only on the set of nonnegative integers up to $n$. 

In the literature, it is also common to represent the confusion matrix (Table~\ref{tab:confusion-raw}) in terms of {\em proportions} rather than in terms of {\em counts}. This is given explicitly in Table~\ref{tab:confusion}, where 
\beqn
\label{eq:discr-cont-relation}
t \equiv \frac{d(k)}{n}, \quad \pi \equiv \frac{n_1}{n}, \quad\mbox{and}\quad 
h(t)\equiv \frac{\hbar(k)}{n}. 
\eeqn
When $n$ is relatively large, it is convenient to think of the function $h(t)$, defined on the interval $[0,1]$, as a continuous object. In fact, we will further assume that it is differentiable almost everywhere. This allows us to use the language of calculus --- i.e., differentiation and integration --- to discuss various concepts. 
The collection of points, $\{(t,h(t)), t \in [0,1]\}$, traces out a so-called {\em hit curve}. For
simplicity, we will refer to $h(t)$ itself as the hit curve as well. Like the ROC curve, the hit curve also is a signature of the underlying  test's effectiveness. Proposition~\ref{prop:basic} below lists a few properties of the hit curve that will be useful later; proofs are given in Appendix~\ref{sec:prop-proof}.

\begin{table}[t]
\centering
\mycap{\label{tab:confusion-raw}%
The confusion matrix based on counts.}
\fbox{%
\begin{tabular}{l|c|c|c}
          & Declared Class-1 & Declared Class-0             & Total \\
\hline
Class-1   & $\hbar(k)$       & $n_1 - \hbar(k)$             & $n_1$ \\ 
Class-0   & $d(k)-\hbar(k)$  & $[n-d(k)] - [n_1-\hbar(k)]$  & $n-n_1$\\
\hline
Total     & $d(k)$           & $n-d(k)$                     & $n$ \\
\end{tabular}}
\end{table}

\begin{table}[t]
\centering
\mycap{\label{tab:confusion}%
The confusion matrix based on proportions.}
\fbox{%
\begin{tabular}{l|c|c|c}
          & Declared Class-1   & Declared Class-0      & Total \\
\hline
Class-1   & $h(t)$   & $\pi - h(t)$         & $\pi$ \\ 
Class-0   & $t-h(t)$ & $(1-t) - [\pi-h(t)]$ & $1-\pi$\\
\hline
Total     & $t$        & $1-t$              & $1$ \\
\end{tabular}}
\end{table}

\begin{proposition}
\label{prop:basic}
Let $h(t)$ be a hit curve (see Table~\ref{tab:confusion}), assumed to be continuous and differentiable almost everywhere. Then, 
\bitem
\item[(a)] $h(0)=0$ and $h(1)=\pi$;
\item[(b)] $0 \leq h'(t) \leq 1$, for all $t$;
\item[(c)] $\int h(t) dh(t) = \pi^2/2$.
\eitem
\end{proposition}

\paragraph{Remark} In what follows, we will {\em not} distinguish between $\hbar$ and $h$. 
Whenever we write $h(k)$, we will be referring to the quantity defined in Table~\ref{tab:confusion-raw} and thinking of it as a discrete function on the set $\{0,1,2,...,n\}$. 
Whenever we write $h(t)$, we will be referring to the quantity defined in Table~\ref{tab:confusion} and thinking of it as a continuous function on $[0,1]$, with $h(t)=h(k)/n$. 

\def\ap{\mbox{AP}}
\def\auc{\mbox{AUC}}
\def\apr{\widetilde{\ap}}
\def\aucr{\widetilde{\auc}}

\subsection{The AUC}
\def\tp{\mbox{TPF}}
\def\fp{\mbox{FPF}}

Suppose we threshold the scores at $x_k$, a level such that $t \times 100\%$ of the subjects are declared to be class-1.
The quantities 
\[
 \tp(t) = \frac{h(t)}{\pi} ~~ \left(\mbox{or} ~~ \tp(k) = \frac{h(k)}{n_1} \right)
\]
and
\[
 \fp(t) = \frac{t-h(t)}{1-\pi} ~~ \left(\mbox{or} ~~ \fp(k) = \frac{d(k)-h(k)}{n-n_1} \right)
\]
are called the {\em true positive fraction} (TPF) and the {\em false positive fraction} (FPF), respectively --- refer to Tables~\ref{tab:confusion-raw}--\ref{tab:confusion}. The ROC curve %(Figure~\ref{fig:roc}) 
refers to the collection of points, $\{(\fp(t), \tp(t)), t \in [0,1]\}$. The AUC is simply its area underneath, defined as
\beqn
 \auc 
&\equiv& \int \tp(t) d [\fp(t)] \label{eq:AUCdefn}.
\eeqn
Using the definitions of $\tp(t)$ and $\fp(t)$ above, it is straight-forward to see that  
\beqn
\auc
&=& \int \frac{h(t)}{\pi} d\left[\frac{t-h(t)}{1-\pi}\right] \notag \\
&=& \frac{1}{\pi(1-\pi)} \int h(t) \left( dt - dh(t)\right) \notag \\
&=& \frac{1}{\pi(1-\pi)} \left[ \int h(t)dt - \int h(t)dh(t) \right] \notag \\
&=& \frac{1}{\pi(1-\pi)} \left[ \int h(t)dt - \frac{\pi^2}{2} \right], \label{eq:auc-auh}
\eeqn
where the final step is due to Proposition~\ref{prop:basic}(c). %Eq.~(\ref{eq:auc-auh}) establishes a direct relationship between the ROC curve and the hit curve. This relationship is known in the finance literature \citep[e.g.,][]{xxx}.

\subsection{The AP}
\def\r{\mbox{Recall}}
\def\p{\mbox{Precision}}

Instead of the TPF and the FPF, the IR community often speaks of the {\em recall} and the {\em precision}.
Again, suppose we threshold the scores at $x_k$, a level such that $t \times 100\%$ of the subjects are declared to be class-1. Then,
\[
 \r(t) = \frac{h(t)}{\pi} ~~ \left(\mbox{or} ~~ \r(k) = \frac{h(k)}{n_1} \right),
\]
which is the same as the TPF, and
\[
 \p(t) = \frac{h(t)}{t} ~~ \left(\mbox{or} ~~ \p(k) = \frac{h(k)}{d(k)} \right).
\]
%again, refer to Tables~\ref{tab:confusion-raw}--\ref{tab:confusion}. 
The average precision (AP) is defined as \citep[e.g.,][]{mz-ap-techrpt} 
\beqn
 \ap &\equiv& \int \p(t) d[\r(t)] \label{eq:APdefn},
\eeqn
which can be thought of as the area under the precision versus recall curve.
Using the definitions of $\r(t)$ and $\p(t)$ above, it is straight-forward to see that 
\beqn
\ap &=& \int \frac{h(t)}{t} \times \frac{dh(t)}{\pi} \notag \\
    &=& \frac{1}{\pi} \int \frac{h(t)}{t} dh(t) \label{eq:AP-comp}.
\eeqn

\subsection{Examples}
\label{sec:extreme-cases}

For those not familiar with either of these concepts, they are often abstract and confusing enough at first sight that a few examples are warranted. For those already comfortable with the ideas, this section can be skipped. 

\subsubsection{A random test} 

If a diagnostic (or screening) test is random, then $h(t)=\pi t$. That is, the true positive rate stays constant at $\pi$, the overall proportion of class-1 subjects. By Eq.~(\ref{eq:auc-auh}), we have
\beqnn
 \auc(\mbox{Random}) 
 = \frac{1}{\pi(1-\pi)} \left[\int \pi t dt - \frac{\pi^2}{2} \right] 
 = 1/2. %\frac{1}{2}.
\eeqnn
By Eq.~(\ref{eq:AP-comp}), we have
\beqnn
 \ap(\mbox{Random})
 = \frac{1}{\pi} \int \frac{\pi t}{t} \pi dt 
 = \pi. 
\eeqnn

\subsubsection{A perfect test} 

If a diagnostic (or screening) test is perfect, then
\[
h(t) = 
\begin{cases}
 t,   & t \leq \pi; \\
 \pi, & t > \pi.
\end{cases}
\]
That is, the true positive rate is 100\% until all class-1 subjects have been identified, after which the true positive rate necessarily stays at zero. 
By Eq.~(\ref{eq:auc-auh}), we have
\beqnn
 \auc(\mbox{Perfect})
 = \frac{1}{\pi(1-\pi)} \left[\int_0^{\pi} t dt + \int_{\pi}^1 \pi dt - \frac{\pi^2}{2} \right] 
 = 1.
\eeqnn
By Eq.~(\ref{eq:AP-comp}), we have
\beqnn
 \ap(\mbox{Perfect})
 = \frac{1}{\pi} \left[ \int_0^{\pi} \frac{t}{t} \times 1 \times dt + \int_{\pi}^1 \frac{\pi}{t}\times 0 \times dt \right]
 = 1. 
\eeqnn

\subsection{Remarks}

Notice that both the AUC and the AP, as we have defined them, are random variables, a point that will become even clearer later in Section~\ref{sec:multinomial}.
Furthermore, according to Eqs.~(\ref{eq:auc-auh}) and (\ref{eq:AP-comp}), both the AUC and the AP are {\em functionals} of the hit curve $h(t)$, a point that we will sometimes emphasize by writing $\auc(h)$ and $\ap(h)$.

\section{Theory}
\label{sec:theory}

In this section, we use a simple, parametric, quasi-concave model for the hit curve to gain some important insight about the AUC and the AP. 

\subsection{The quasi-concave model}

\begin{figure}[p]
\centering
\includegraphics[width=0.9\textwidth]{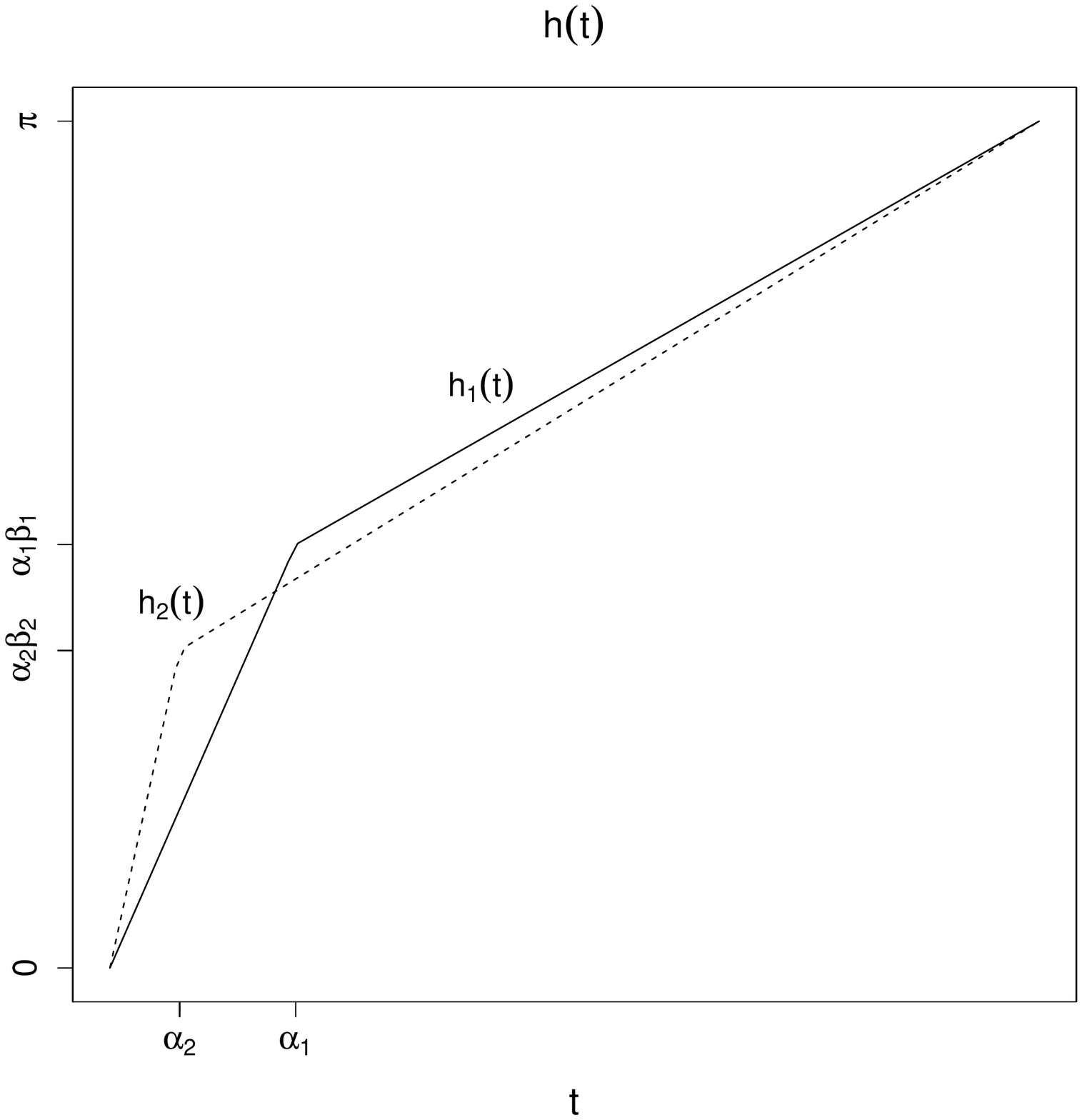}
\mycap{\label{fig:exB}%
Two quasi-concave hit curves, $h_1(t)$ and $h_2(t)$, parameterized respectively by $(\alpha_1, \beta_1)$ and $(\alpha_2, \beta_2)$. In this illustration, the parameters $\alpha_1, \alpha_2, \beta_1, \beta_2$ are configured so that $\auc(h_1)=\auc(h_2)$ --- see Theorem~\ref{thm:equal-auc-pwlinear}. But since $\beta_2 > \beta_1$, Theorem~\ref{thm:apauc-pwlinear} implies that $\ap(h_2) > \ap(h_1)$. In practice, the initial part of the hit curve is often more relevant, which makes the AP a more useful evaluation metric.}
\end{figure}

Consider a quasi-concave hit curve (Figure~\ref{fig:exB}), parameterized as follows:
\beqn
\label{eq:quasi-concave}
h(t) = 
\begin{cases}
\beta t, & t \in [0, \alpha]; \\
 & \\
\frac{\displaystyle \pi-\alpha\beta}{\displaystyle 1-\alpha}\left(t-\alpha\right)+\alpha\beta, & t \in (\alpha,1].
\end{cases}
\eeqn
There are two parameters: $\beta \in [\pi, 1]$ is the {\em initial true positive rate} of the underlying test, and $\alpha \in [0, \pi/\beta]$ is the {\em change point} at which the test's true positive rate drops. The requirement $\beta \geq \pi$ ensures that the model is describing a test that is as least as good as random; worse-than-random procedures are not interesting and practically irrelevant. The requirement $\beta \leq 1$ is due to Proposition~\ref{prop:basic}(b). And, finally, the requirement $\alpha \leq \pi/\beta$ is because, with a true positive rate of $\beta$, all class-1 subjects will have been identified by $t=\pi/\beta$, leaving no more for $t > \pi/\beta$.

\subsection{Momentum and stamina}
\label{sec:newdef}

Despite its overwhelming simplicity, there are good reasons why the quasi-concave model (\ref{eq:quasi-concave}) is useful. 

First, hit curves are typically concave, reflecting the fact that the true positive rate typically drops as the scores decrease --- that is, higher-ranked subjects are more likely than lower-ranked ones to belong to class-1. Eq.~(\ref{eq:quasi-concave}) is arguably the simplest approximation possible to any concave function. When $\beta > \pi$, it is easy to see that the slope of the second segment, $(\pi-\alpha\beta)/(1-\alpha)$, is smaller than that of the first segment, $\beta$, which is why we use the name ``quasi-concave''. This idea of using a quasi-concave function to approximate a concave function is inspired by \citet{laibson}, who used ``quasi-hyperbolic'' discount functions to study time-inconsistent intertemporal choices in behavioral economics. 

Second, the two parameters, $\alpha$ and $\beta$, each capture an essential feature of the underlying diagnostic (or screening) test:

\bitem

\item[$\alpha$:] As the change point at which the true positive rate drops, this parameter measures the {\em stamina} of the test --- how long can the initial, relatively high true positive rate ``last''?

\item[$\beta$:] As the initial true positive rate, this parameter measure the {\em momentum} of the test --- performing at its best level, how fast can the test identify the subjects that it is supposed to identify?

\eitem

\subsection{Momentum-stamina trade-off}
\label{sec:tradeoff}

For a quasi-concave hit curve, $h(t)$, given by Eq.~(\ref{eq:quasi-concave}), it is easy to see that
\beqn
 \int h(t) dt 
= \frac{\beta\alpha^2}{2} + \frac{(\pi+\beta\alpha)(1-\alpha)}{2}
= \frac{\pi+\beta\alpha-\pi \alpha}{2}
= \frac{\pi+(\beta-\pi)\alpha}{2}. \label{eq:fromAUH}
\eeqn
Then, by Eq.~(\ref{eq:auc-auh}), 
\beqn
\auc 
&=& \frac{1}{2\pi(1-\pi)} \left[\pi + (\beta-\pi)\alpha - \pi^2 \right] \notag \\
&=& \frac{1}{2\pi(1-\pi)}[(\beta-\pi)\alpha] + \frac{1}{2}, \label{eq:auc-stuff}
\eeqn
which immediately implies Theorem~\ref{thm:equal-auc-pwlinear} below.
\begin{theorem}
\label{thm:equal-auc-pwlinear} 
If two hit curves, $h_1(t)$ and $h_2(t)$, both belong to the quasi-concave family (\ref{eq:quasi-concave}), and are parameterized respectively by $(\alpha_1, \beta_1)$ and $(\alpha_2, \beta_2)$, then $\auc(h_1) = \auc(h_2)$ if and only if 
\beqn
%\label{eq:iff-equal-auc-pwlinear}
 (\beta_1 - \pi) \alpha_1 = (\beta_2 - \pi) \alpha_2. 
\quad\square
\eeqn
\end{theorem}
Theorem~\ref{thm:equal-auc-pwlinear} explains that two diagnostic (or screening) tests $h_1$ and $h_2$ can have the same AUC for {\em different} reasons. The trivial case is when both $\alpha_1 = \alpha_2$ and $\beta_1 = \beta_2$; this is when $h_1$ and $h_2$ are truly identical --- same stamina and same momentum. However, if $\beta_1 < \beta_2$, then Theorem~\ref{thm:equal-auc-pwlinear} implies that we must necessarily have $\alpha_1 > \alpha_2$, and vice versa. That is, on the AUC-scale, mediocre momentum can be compensated by greater stamina, and vice versa. This provides a mathematically explicit explanation for the reason behind criticism (C1), namely why two qualitatively different tests can end up having similar AUC values.

\subsection{AP versus AUC}
\label{sec:AP-extra-momentum}

What about the AP? If $h(t)$ is a quasi-concave hit curve given by Eq.~(\ref{eq:quasi-concave}), then 
\[
dh(t) = 
\begin{cases}
\beta dt, & t \in [0, \alpha]; \\
 & \\
\frac{\displaystyle \pi-\alpha\beta}{\displaystyle 1-\alpha} dt, & t \in (\alpha,1].
\end{cases}
\]
So, 
\beqnn
\ap 
&=& \frac{1}{\pi} \int \frac{h(t)}{t} dh(t) \\
&=& \frac{1}{\pi}
\left[
 \int_0^{\alpha} \beta^2 dt + 
 \int_{\alpha}^{1} \left(\frac{\pi-\alpha\beta}{1-\alpha}\right)^2 + 
   \frac{\left(\frac{\pi-\alpha\beta}{1-\alpha}\right)\alpha\beta-
         \left(\frac{\pi-\alpha\beta}{1-\alpha}\right)^2 \alpha}{t} dt
\right]\\
&=& \frac{1}{\pi}
\left[
 \beta^2 \alpha + \frac{(\pi-\alpha\beta)^2}{1-\alpha} -
 \left(\frac{\pi-\alpha\beta}{1-\alpha}\right)\frac{(\beta-\pi)\alpha}{1-\alpha} \log \alpha
\right].
\eeqnn
Using the Taylor approximation that $\log(\alpha) \approx \alpha-1$, the expression 
above can be simplified to
\beqn
\ap 
\approx \frac{\beta^2 \alpha}{\pi} + \pi - \alpha\beta 
= \frac{\beta}{\pi} \left[(\beta-\pi)\alpha\right] + \pi. \label{eq:fromAP}
\eeqn
Recall from Section~\ref{sec:extreme-cases} that 
$\ap(\mbox{Random})=\pi$, $\ap(\mbox{Perfect})=1$, 
$\auc(\mbox{Random})=1/2$, and $\auc(\mbox{Perfect})=1$. 
We can rescale the AP and the AUC to both lie between $0$ and $1$:
\beqn
\label{eq:rescaledAP}
\apr \equiv \frac{\ap-\pi}{1-\pi}
\eeqn
and
\beqn
\label{eq:rescaledAUC}
\aucr \equiv \frac{\auc-1/2}{1-1/2} = 2\auc - 1.
\eeqn
Then, 
\[
\aucr = \frac{1}{\pi(1-\pi)} \times \left[ (\beta-\pi)\alpha \right]
\]
by Eq.~(\ref{eq:auc-stuff}) and Eq.~(\ref{eq:rescaledAUC}), while 
\[
\apr \approx 
\frac{1}{\pi(1-\pi)} \times \beta \times \left[ (\beta-\pi)\alpha \right]
\]
by Eq.~(\ref{eq:fromAP}) and Eq.~(\ref{eq:rescaledAP}). These results establish Theorem~\ref{thm:apauc-pwlinear} below.
\begin{theorem}
\label{thm:apauc-pwlinear}
If a hit curve, $h(t)$, belongs to the quasi-concave family (\ref{eq:quasi-concave}), then
\beqn
\apr(h) \approx \beta \times \aucr(h). \label{eq:apauc-pwlinear}
\quad\square
\eeqn
\end{theorem}
Theorem~\ref{thm:apauc-pwlinear} suggests that, if two diagnostic (or screening) tests have the same AUC, then the AP will ``reward extra points'' to the one with the larger momentum (larger $\beta$). Since momentum is the initial true positive rate, this means the AP places more emphasis on the initial part of the ROC curve. Hence, the AP can be seen as being based upon the AUC but having incorporated a self-correcting factor in response to criticism (C2).

\subsection{A simple simulation}
\label{sec:sim}

The theoretical insights derived in the previous sections are based on using a quasi-concave model for the hit curve. Here, we report a simple simulation study to investigate the applicability of Theorem~\ref{thm:apauc-pwlinear} to general hit curves, i.e., those outside the quasi-concave family, Eq.~(\ref{eq:quasi-concave}).

In the literature \citep[see, e.g.,][]{rocbk}, it is customary to simulate diagnostic/screening tests in the following way: Without loss of generality, scores given by the test to subjects in class-0 are simulated from $f_0(x) \sim \normal{0}{1}$, and those given to subjects from class-1 are simulated from $f_1(x) \sim \normal{\Delta}{1}$ for some $\Delta > 0$. How well the scores rank the subjects (in terms of the AP and/or the AUC) is then studied. The parameter, $\Delta$, controls the strength of the simulated test --- a large $\Delta$ means the test tends to give much higher scores to subjects in class-1 than to those in class-0; it is thus a more powerful test.

Given $(n, \pi, \Delta)$, we first generated $n\pi$ scores from $\normal{\Delta,1}$ and $n(1-\pi)$ scores from $\normal{0}{1}$. Using these scores, we plotted the resulting hit curve, $h(t)$, as well as computed $\ap(h)$ and $\auc(h)$. Then, we computed
\beqn
 \widehat{\beta} = \frac{\apr(h)}{\aucr(h)} =
 \frac{(\ap(h)-\pi)/(1-\pi)}{2\auc(h)-1}, \label{eq:betahat}
\eeqn
and plotted the line $f(t) = \widehat{\beta}t$ on top of $h(t)$. This procedure was easily repeated for many combinations of $(n, \pi, \Delta)$.

\begin{figure}[p]
\centering
\includegraphics[width=0.9\textwidth]{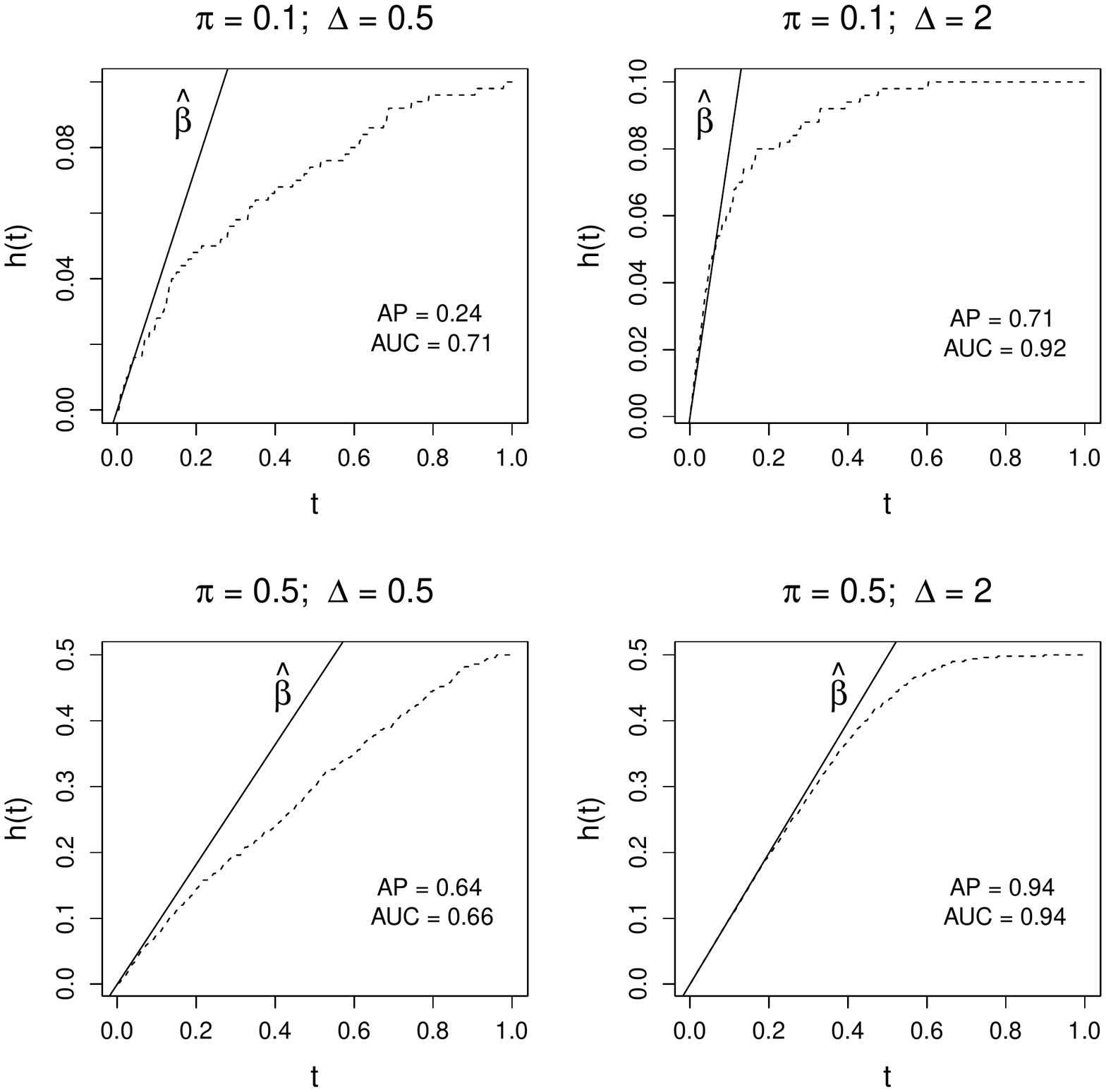}
\mycap{\label{fig:showapprox}%
Illustration of $\widehat{\beta}$ --- Eq.~(\ref{eq:betahat}) --- under four simulated scenarios, representing different levels of prevalence, high ($\pi=0.5$) versus low ($\pi=0.1$), and different strengths of the diagnostic/screening test, strong ($\Delta = 2$) versus weak ($\Delta = 0.5$). Dashed lines are hit curves. Solid lines are the function, $f(t)=\widehat{\beta}t$. A total of $n\pi$ scores are simulated from $\normal{\Delta}{1}$, and the remaining $n(1-\pi)$ scores are simulated from $\normal{0}{1}$, with $n=500$.}
\end{figure}

Fixing $n=500$, Figure~\ref{fig:showapprox} shows four representative scenarios, generated by all combinations of 
$\pi=0.1$ (low prevalence) or $\pi=0.5$ (high prevalence), and $\Delta=0.5$ (weak test) or $\Delta=2$ (strong test). Our results, including those not shown in Figure~\ref{fig:showapprox}, suggest that $\widehat{\beta}$, as given by equation~(\ref{eq:betahat}), is a good approximation of the initial true positive rate. This, in turn, verifies that the approximate relationship (\ref{eq:apauc-pwlinear}) established by Theorem~\ref{thm:apauc-pwlinear} is still useful even if the hit curve $h(t)$ does {\em not} belong to the quasi-concave family (\ref{eq:quasi-concave}).

\section{Practice}
\label{sec:multinomial}

While conceptually it is convenient to think of $h(t)$ as a continuous function, in practice we are often faced with its discrete cousin, $h(k)$. In this section, we first describe the typical set-up (Section~\ref{sec:setup}). In Section~\ref{sec:discrete-theory}, we derive explicit expressions for the AUC and the AP under this set-up. These expressions not only show how the AUC and the AP can be computed in practice, but they also confirm, from a slightly different point of view, our earlier result that the AP places more emphasis on initial true positives. To actually use the AP as a performance metric in practice, we need not only the AP, but also its standard error. In Section~\ref{sec:variance}, we derive the asymptotic variance for the AP by considering the set-up given in Section~\ref{sec:setup}. 
%, so that we can assess whether two tests are statistically different in terms of the AP.

\begin{table}[h]
\centering
\mycap{\label{tab:K-partition}%
A diagnostic (or screening) test partitions $n$ subjects into $K$ groups ($K$ distinct scores). The broken bars (\textbrokenbar) illustrates the case where all those with scores $\geq x_k$ (left) are declared to be class-1, while all those with scores $< x_k$ (right) are declared to be class-0.}
\fbox{
\begin{tabular}{l|ccccccccc|c}
Score     & $x_1$ & $>$ & $x_2$     &  $>\cdots>$  & $x_k$   & $>$ & $x_{k+1}$ &  $>\cdots>$  & $x_K$ &  \\ 
Partition & $R_1$ &     & $R_2$     &   $\cdots$   & $R_k$   & \textbrokenbar    & $R_{k+1}$ &  $\cdots$    & $R_K$  & Total \\
\hline
Class-1  & $Z_1$       & & $Z_2$       & $\cdots$ & $Z_k$       & \textbrokenbar & $Z_{k+1}$       & $\cdots$ & $Z_K$       & $n_1$ \\
Class-0  & $\bar{Z}_1$ & & $\bar{Z}_2$ & $\cdots$ & $\bar{Z}_k$ & \textbrokenbar & $\bar{Z}_{k+1}$ & $\cdots$ & $\bar{Z}_K$ & $n_0$ \\
 \hline
Total    & $S_1$       & & $S_2$       & $\cdots$ & $S_k$       & \textbrokenbar & $S_{k+1}$       & $\cdots$ & $S_K$       & $n$ \\
\end{tabular}}
\end{table}

\subsection{A typical set-up}
\label{sec:setup}

In general, suppose a diagnostic (or screening) test gives $K$ distinct scores for a total of $n$ subjects, with $K \leq n$. If $K < n$, it means that some subjects' scores are tied. The case of ``no ties'' simply corresponds to the special case of $K=n$. With $K$ distinct scores, the subjects are partitioned into $K$ groups. Within each group, some may belong to class-1 and others may belong to class-0, but the test cannot distinguish them. We will use $R_1$ to denote the set of all subjects receiving the top score, $R_2$ to denote the set of all subjects receiving the next top score, and so on for $R_3, ..., R_K$. Furthermore, let
\beqnn
 S_k &=& \mbox{total number of subjects in $R_k$}, \\
 Z_k &=& \mbox{total number of class-1 subjects in $R_k$}, \\
 \bar{Z}_k &=& \mbox{total number of class-0 subjects in $R_k$}.
\eeqnn
Table~\ref{tab:K-partition} summarizes the set-up and the notations we have just introduced.

\subsection{AP versus AUC, again}
\label{sec:discrete-theory}

Under the typical set-up (Table~\ref{tab:K-partition}), if we threshold the scores at $x_k$, then all those in partitions $R_1, R_2, ..., R_k$ will be declared class-1 and the rest, declared class-0. Therefore, we have
\beqnn
 d(k) = \sum_{k' \leq k} S_{k'}
 \quad\mbox{and}\quad
 h(k) = \sum_{k' \leq k} Z_{k'}. 
% p(k) &=& \frac{h(k)}{d(k)}  = \frac{\sum_{k' \leq k} Z_{k'}}{\sum_{k' \leq k} S_{k'}}, \\
% r(k) &=& \frac{h(k)}{n_1} = \frac{\sum_{k' \leq k} Z_{k'}}{n_1}.
\eeqnn
As a result, Eq.~(\ref{eq:auc-auh}) becomes
\beqnn
\mbox{AUC} 
&=& \frac{1}{(n_1/n)(n_0/n)} 
 \left\{ 
 \sum_{k=1}^K \left[\frac{h(k)}{n}\right] \left[\frac{\Delta d(k)}{n}\right] - \frac{1}{2}\left(\frac{n_1}{n}\right)^2
 \right\} \\
&=& \frac{n}{n_0} 
 \left\{ 
 \sum_{k=1}^K \left[\frac{h(k)}{n_1}\right] \left[ \frac{\Delta d(k)}{n} \right] 
 \right\} - \frac{1}{2} \left(\frac{n_1}{n_0}\right), 
\eeqnn
where the term inside the curly brackets can be simplified further as follows:
\beqn 
& & \sum_{k=1}^K \left[\frac{h(k)}{n_1}\right] \left[ \frac{\Delta d(k)}{n} \right] \notag \\
&=& 
\left[\frac{Z_1}{n_1}\right]\left[\frac{S_1}{n}\right] +
\left[\frac{Z_1+Z_2}{n_1}\right]\left[\frac{S_2}{n}\right] + ... +
\left[\frac{Z_1+Z_2+...+Z_K}{n_1}\right]\left[\frac{S_K}{n}\right] \notag \\
&=& 
\underbrace{\left[\frac{S_1+S_2+...+S_K}{n}\right]}_{w_1^{'}}\left[\frac{Z_1}{n_1}\right] +
\underbrace{\left[\frac{S_2+...+S_K}{n}\right]}_{w_2^{'}}\left[\frac{Z_2}{n_1}\right] + ... +
\underbrace{\left[\frac{S_K}{n}\right]}_{w_K^{'}}\left[\frac{Z_K}{n_1}\right] \label{eq:AUC-weights}\\
&=& \sum_{k=1} w_k^{'} \left[ \frac{Z_k}{n_1} \right] \label{eq:AUC-multinomial}.
\eeqn
Likewise, Eq.~(\ref{eq:AP-comp}) becomes
\beqn
\mbox{AP} 
%&=& \frac{1}{n_1} \sum_{j} \left[\frac{h(t_j)}{t_j}\right] \Delta h(t_j)  \\
&=& \frac{1}{(n_1/n)} \sum_{k=1}^K \left[\frac{h(k)}{d(k)}\right] \left[\frac{\Delta h(k)}{n}\right] \notag \\
&=& \sum_{k=1}^K \left[\frac{h(k)}{d(k)}\right] \left[\frac{\Delta h(k)}{n_1}\right] \notag  \\
&=& 
\underbrace{\left[\frac{Z_1}{S_1}\right]}_{w_1}\left[\frac{Z_1}{n_1}\right] +
\underbrace{\left[\frac{Z_1+Z_2}{S_1+S_2}\right]}_{w_2}\left[\frac{Z_2}{n_1}\right] + ... +
\underbrace{\left[\frac{Z_1+Z_2+...+Z_K}{S_1+S_2+...+S_K}\right]}_{w_K}\left[\frac{Z_K}{n_1}\right] \label{eq:AP-discrete} \\
&=& \sum_{k=1} w_k \left[ \frac{Z_k}{n_1} \right]. \label{eq:AP-multinomial}
\eeqn
Eqs.~(\ref{eq:AUC-weights}) and (\ref{eq:AP-discrete}) give convenient and explicit expressions for how the AUC and the AP are calculated in practice. They also reveal that both the AUC and the AP can be expressed as weighted averages of $Z_1, Z_2, ..., Z_K$, except that they use different weights: $w_k^{'}$ for the AUC and $w_k$ for the AP. 

The difference between $w_k^{'}$ and $w_k$ can most clearly be seen in the case of ``no ties'', i.e., $K=n$. Under such circumstances, each $R_k$ contains just one subject, so $S_k=1$ for all $k$, and each $Z_k$ is either zero %(indicating it's a false positive) 
or one. %(indicating it's a true positive). 
Then, it is easy to see from Eqs.~(\ref{eq:AUC-weights})-(\ref{eq:AUC-multinomial}) and Eqs.~(\ref{eq:AP-discrete})-(\ref{eq:AP-multinomial}) that
\[
 w_k^{'} = \frac{n-d(k)}{n} = \frac{n-k+1}{n}
 \quad\mbox{and}\quad
 w_k = \frac{h(k)}{k} = \frac{(\mbox{number of true positives up to $k$})}{k}. %=\frac{h(k)}{k}.
\]
The main difference, therefore, is that the weights used by the AUC are {\em independent} of, whereas the ones used by the AP are {\em adaptive} to, the test itself. Suppose $Z_{k_1}=1$ and $Z_{k_2}=1$ both belong to class-1, where $k_1 < k_2$. When calculating the AUC for two different tests, A and B, these true positives will each receive a fixed weight, $(n-k_1-1)/n$ and $(n-k_2-1)/n$, respectively. For the AP, the weights they receive will depend on the strength of the test itself; in particular, if test A identified more class-1 subjects before $k_1$ than did test B, the relative weight on $Z_{k_1}$ would be bigger for test A than for test B. This shows, from a different point of view, that the AP places more emphasis on early true positives than does the AUC.

\subsection{Asymptotic variance of the AP}
\label{sec:variance}
\newcommand{\Perp}{\perp \! \! \! \perp}
\newcommand{\wh}{\widehat}

In order to use the AP as an evaluation metric, we need to obtain a formula for its variance, so that its standard error can be computed in practice. 
To derive such a formula, notice that the data in Table~\ref{tab:K-partition} follow the ensuing distributions:
\beqn
(Z_1, Z_2, ..., Z_K)|n_1 &\sim& \mbox{multinomial}(n_1; p_1, p_2, ..., p_K), \label{eq:mult-p}\\
(\bar{Z}_1, \bar{Z}_2, ..., \bar{Z}_K)|n_1 &\sim& \mbox{multinomial}(n-n_1; q_1, q_2, ..., q_K), \label{eq:mult-q}\\
n_1 &\sim& \mbox{binomial}(n, \pi), \label{eq:mult-pi} 
\eeqn
where
\beqnn
 p_k = \int_{R_k} f_1(x) dx, \quad
 q_k = \int_{R_k} f_0(x) dx,
\eeqnn 
and $f_j(x)$ is the distribution of the scores in class-$j$ ($j=0,1$). In addition, 
\[
(Z_1, Z_2, ..., Z_K) \Perp (\bar{Z}_1, \bar{Z}_2, ..., \bar{Z}_K)|n_1. 
\]
Therefore, the log-likelihood function (aside from a constant) is given by
\beqn
\label{eq:loglik}
\ell(\myvec{p},\myvec{q},\pi) = 
\sum_{k=1}^K z_k \log p_k + \sum_{k=1}^K \bar{z}_k \log q_k + \left[n_1\log\pi + (n-n_1)\log(1-\pi)\right],
\eeqn
where $\myvec{p} \equiv (p_1, \cdots, p_K)^{\T}, \myvec{q} =(q_1, \cdots, q_K)^{\T}$, 
\beqnn
\sum_{k=1}^K p_k=1, \quad \sum_{k=1}^K q_k=1, \quad
\sum_{k=1}^K z_k=n_1, \quad\mbox{and}\quad \sum_{k=1}^K \bar{z}_k=n-n_1.
\eeqnn
Let $(\wh{\myvec{p}}, \wh{\myvec{q}}, \wh{\pi})$ denote the MLEs of $(\myvec{p}, \myvec{q}, \pi)$. 
Then, by classical theory \citep[e.g.,][]{coxhinkley}, the asymptotic variance of $(\wh{\myvec{p}}, \wh{\myvec{q}}, \wh{\pi})$ is simply $\myvec{J}^{-1}$, where $\myvec{J}$ is the Fisher information matrix associated with the log-likelihood function, Eq.~(\ref{eq:loglik}). Thus, if we can express the AP as a function of $(\wh{\myvec{p}}, \wh{\myvec{q}}, \wh{\pi})$, say $\ap = g(\wh{\myvec{p}}, \wh{\myvec{q}}, \wh{\pi})$, then we easily can estimate its asymptotic variance using the delta method, i.e.,
\beqn
\label{eq:AP-var}
\widehat{\var}(\ap) = \left(\nabla g\right)^{\T} \widehat{\myvec{J}}^{-1} \left(\nabla g\right),
\eeqn
where $\widehat{\myvec{J}}$ denotes the {\em observed} Fisher information matrix.
Indeed, this can be done based on Eq.~(\ref{eq:AP-discrete}) and the fact that
\[
\wh{p}_k = \frac{Z_k}{n_1} , \quad
\wh{q}_k = \frac{\bar{Z}_k}{n-n_1}, \quad\mbox{and}\quad
\wh{\pi} = \frac{n_1}{n}.
\]
In particular, it is easy to see that
\beqn
\label{eq:AP-g}
\ap = g(\wh{\myvec{p}}, \wh{\myvec{q}}, \wh{\pi}) = \sum_{k=1}^K \left[ \wh{p}_k 
\left(
\frac{\displaystyle \wh{\pi}\sum_{k' \leq k} \wh{p}_{k'}}
     {\displaystyle \wh{\pi}\sum_{k' \leq k} \wh{p}_{k'}+(1-\wh{\pi})\sum_{k' \leq k} \wh{q}_{k'}}
\right) \right].
\eeqn
Let $\gvec{\theta} \equiv (\myvec{p}, \myvec{q}, \pi)$ and $\wh{\gvec{\theta}} \equiv (\wh{\myvec{p}}, \wh{\myvec{q}}, \wh{\pi})$. The calculations of
\beqnn
\widehat{\myvec{J}} = \widehat{\E}\left[-\frac{\partial \ell^2}{\partial \gvec{\theta} \partial \gvec{\theta}^{\T}}\right] 
\quad\mbox{and}\quad 
\nabla g = \frac{\partial g}{\partial \wh{\gvec{\theta}}}
\eeqnn
are straight-forward; details are given in Appendix~\ref{sec:APvar-details}.

\section{Examples}
\label{sec:example}

We now provide two illustrative examples.

\subsection{Mass spectrometry data for prostate cancer}
\label{sec:prostate}

Our first example concerns protein biomarkers for prostate cancer. By analyzing serum samples obtained from the Virginia Prostate Center Tissue and Body Fluid Bank, \citet{AP-prostate-example} identified 779 potential biomarkers using a technology called ``surface-enhanced laser desorption/ionization time-of-flight mass spectrometry'' \citep{seldi-tof-ms}. \citet{wang-pAUC} used this data set to illustrate the partial AUC (see Section~\ref{sec:intro}). We focused on the $n_1=83$ patients with late-stage prostate cancer (class 1) and the $n_0=82$ normal individuals (class 0) in the data set, although the original data set also included patients with early-stage cancer and patients with benign prostate hyperplasia.

\begin{figure}[th]
\centering
\includegraphics[width=0.9\textwidth]{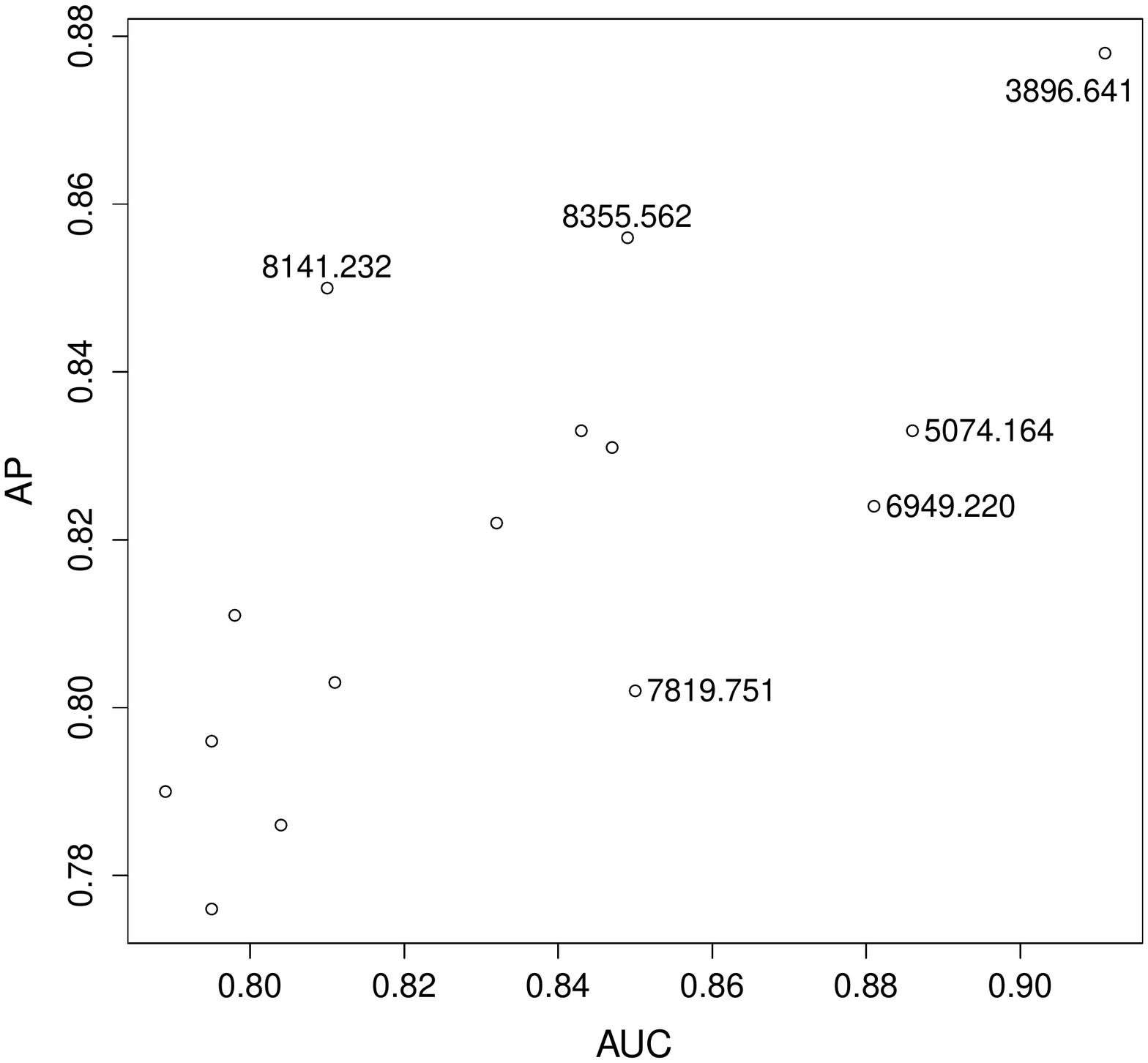}
\mycap{\label{fig:prostateAPAUC}%
Prostate cancer example (Section~\ref{sec:prostate}). Top 15 biomarkers according to the AP. Biomarkers are {\em not} labelled unless they are explicitly mentioned in the text.}
\end{figure}

Figure~\ref{fig:prostateAPAUC} shows the AP versus the AUC for the top 15 biomarkers as ranked by the AP. We can see clearly that, while some biomarkers (e.g., $3896.614$) are ranked similarly on both scales, others (e.g., $8355.562$, $8141.232$, $5074.164$, $6949.220$, $7819.751$) are ranked very differently. For example, according to both the AP and the AUC, $3896.641$ is a top biomarker. On the other hand, according to the AUC, there is little difference between $8355.562$ and $7819.751$ whereas, according to the AP, $8355.562$ is more powerful.

\begin{figure}[th]
\centering
\subfigure[Pair A: ($8355.562$, $7819.751$).]{\includegraphics[width=0.475\textwidth]{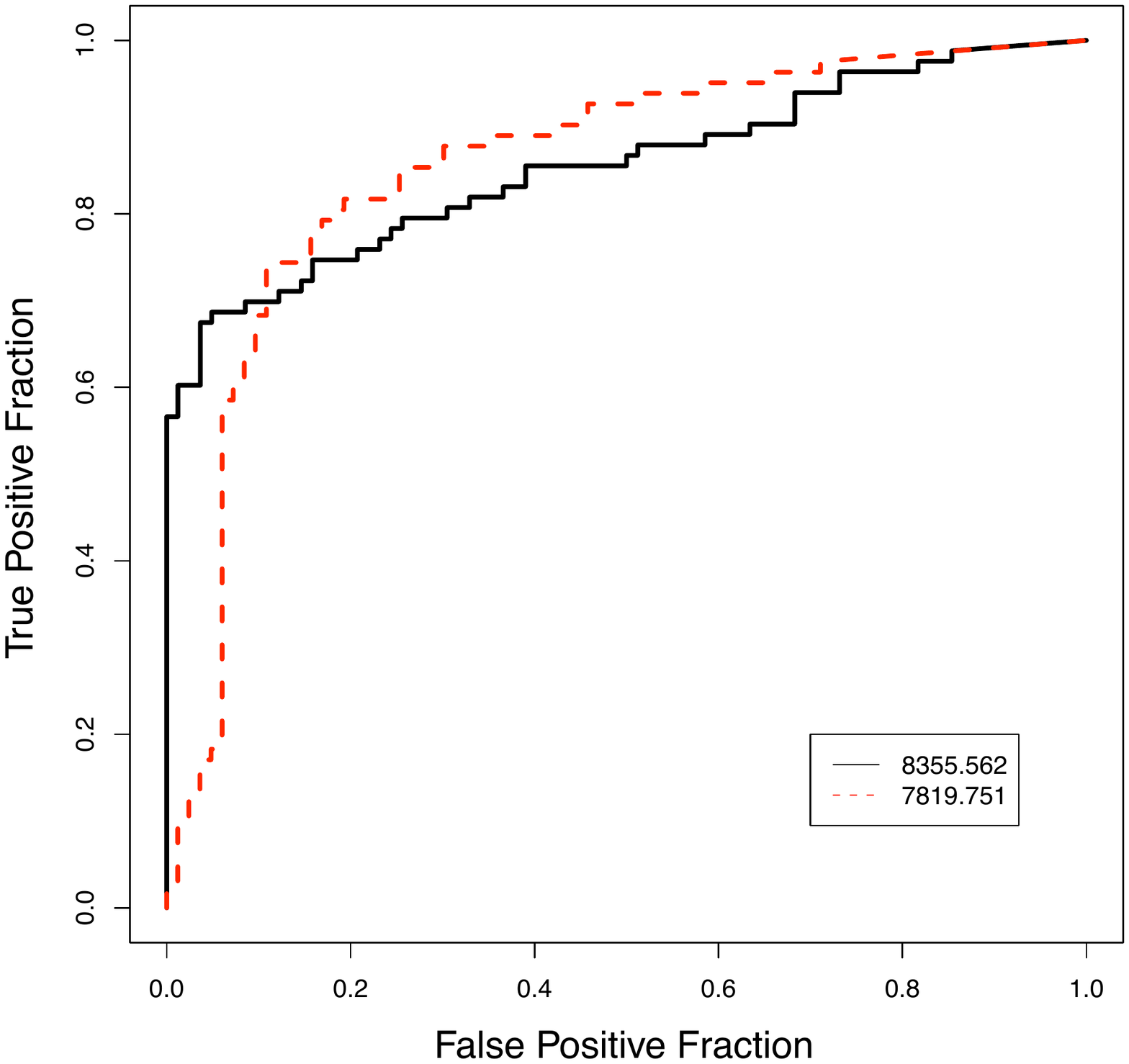}}
\subfigure[Pair B: ($8141.232$, $5074.164$).]{\includegraphics[width=0.475\textwidth]{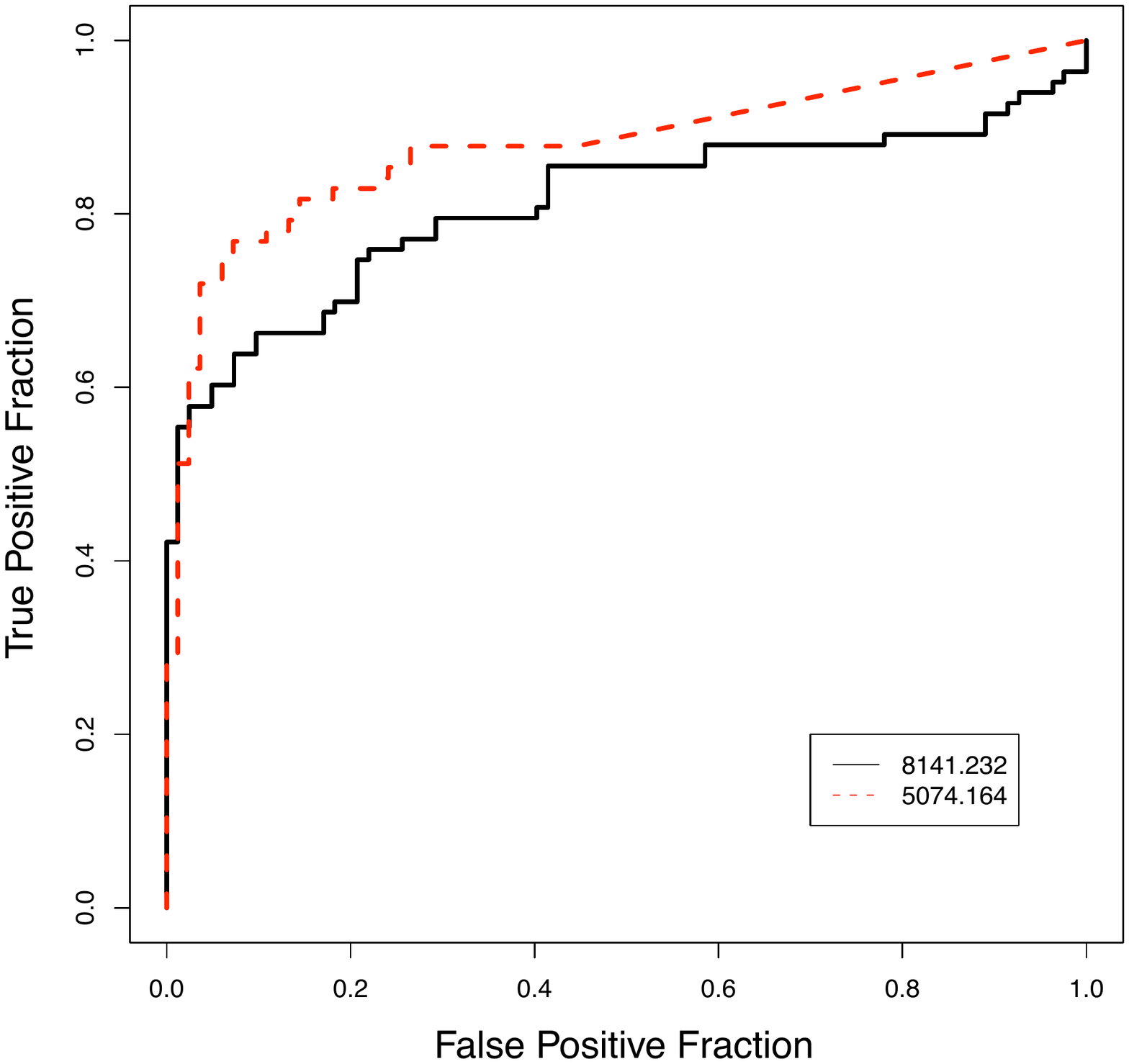}}
\mycap{\label{fig:prostateROC}%
Prostate cancer example (Section~\ref{sec:prostate}). Comparison of ROC curves for biomarkers that are rankly differently by the AP and by the AUC. Pair A $(8355.562, 7819.751)$ scored similarly on the AUC-scale but very differently on the AP-scale. Pair B $(8141.232, 5074.164)$ scored more or less similarly on the AP-scale but very differently on the AUC-scale.}
\end{figure}

Figure~\ref{fig:prostateROC} compares the ROC curves of two selected pairs of biomarkers: A ($8355.562$ and $7819.751$), which scored quite similarly on the AUC-scale but very differently on the AP-scale; and B ($8141.232$ and $5074.164$), which scored more or less similarly on the AP-scale but very differently on the AUC-scale. We clearly can see from this figure that the two biomarkers in pair A have qualitatively different ROC curves, yet their AUC-values are very similar. This is an unmistakable case of the momentum-stamina tradeoff that we discussed in Section~\ref{sec:tradeoff} --- even though the areas under their ROC curves are obviously very similar, the biomarker $8355.562$ also can be recognized from its ROC curve to have a much larger momentum, which is why it scored much higher on the AP-scale, since the AP awards extra points to momentum (Section~\ref{sec:AP-extra-momentum}). For the two biomarkers in pair B, one immediately can discern from Figure~\ref{fig:prostateROC} that $5074.164$ has a much larger area under its ROC curve (i.e., larger AUC), yet their AP-values are more or less the same --- in fact, the AP of $8141.232$ is slightly higher than that of $5074.164$. Again, this is due to the biomarker $8141.232$ having a slightly larger momentum, a fact also noticeable from the figure.  

Table~\ref{tab:wang-rslt} contains various standard error estimates for the top $15$ biomarkers displayed in Figure~\ref{fig:prostateAPAUC}. Here, we can see that the asymptotic estimates (Section~\ref{sec:variance}) do, in fact, agree closely with standard bootstrap estimates \citep{bootstrap}. To obtain the parametric bootstrap estimates, we first estimated the multinomial models (\ref{eq:mult-p})--(\ref{eq:mult-pi}), and then used the estimated parameters ($\wh{p}_k, \wh{q}_k, \wh{\pi}$) to generate $B=5,000$ multinomial samples. For the nonparametric bootstrap, we simply drew $B=5,000$ bootstrap samples, each by resampling from the original data set with replacement.
 
\begin{table}[p]
\centering
\mycap{Prostate cancer example (Section~\ref{sec:prostate}). The top 15 biomarkers according to the AP. %
``P-Bootstrap'' = parametric bootstrap. ``NP-Bootstrap'' = nonparametric bootstrap.}
\label{tab:wang-rslt}
\fbox{%
\begin{tabular}{lr|c|ccc}
 &      &    & \multicolumn{3}{c}{Standard Error of AP}  \\
\multicolumn{2}{l|}{Biomarker}   & AP & Asymptotic & P-Bootstrap & NP-Bootstrap               \\
\hline
1  & 3896.641  & 0.878  & 0.0345  & 0.0344  & 0.0344  \\
2  & 8355.562  & 0.856  & 0.0336  & 0.0339  & 0.0340  \\
3  & 8141.232  & 0.850  & 0.0319  & 0.0324  & 0.0321  \\
4  & 8295.641  & 0.833  & 0.0328  & 0.0327  & 0.0327  \\
5  & 5074.164  & 0.833  & 0.0403  & 0.0405  & 0.0403  \\
6  & 4071.184  & 0.831  & 0.0368  & 0.0364  & 0.0366  \\
7  & 6949.220  & 0.824  & 0.0414  & 0.0415  & 0.0413  \\
8  & 9149.121  & 0.822  & 0.0378  & 0.0380  & 0.0378  \\
9  & 5914.398  & 0.811  & 0.0355  & 0.0345  & 0.0355  \\
10 & 28142.463  & 0.803  & 0.0395  & 0.0394  & 0.0402  \\
11 & 7819.751  & 0.802  & 0.0424  & 0.0427  & 0.0423  \\
12 & 7195.206  & 0.796  & 0.0317  & 0.0316  & 0.0318  \\
13 & 16264.029  & 0.790  & 0.0320  & 0.0321  & 0.0319  \\
14 & 7775.625  & 0.786  & 0.0441  & 0.0451  & 0.0442  \\
15 & 8544.842  & 0.776  & 0.0388  & 0.0389  & 0.0390  
\end{tabular}}
\end{table}

\begin{table}[p]
\centering
\mycap{Breast cancer example (Section~\ref{sec:breast}). Film versus digital mammography. %
``P-Bootstrap'' = parametric bootstrap. ``NP-Bootstrap'' = nonparametric bootstrap.}
\label{tab:breast-rslt}
\fbox{%
\begin{tabular}{l|c|c|ccc}
Mammography       &     &    & \multicolumn{3}{c}{Standard Error of AP}  \\
Type   & AUC & AP & Asymptotic & P-Bootstrap & NP-Bootstrap               \\
\hline
Digital  & 0.753 & 0.144  & 0.0197  & 0.0197  & 0.0194  \\
Film     & 0.735 & 0.166  & 0.0219  & 0.0216  & 0.0215  
\end{tabular}}
\end{table}

\subsection{Mammography data for breast cancer}
\label{sec:breast}

Our second example concerns the Digital Mammographic Imaging Screening Trial \citep[DMIST;][]{AP-breast-example}, comparing digital versus film mammography for breast cancer screening. 
Over 42,000 women were enrolled in the trial and underwent both digital and film mammography.  
Using a seven-point malignancy scale, each pair of mammograms were rated separately by two independent radiologists. 
At 15-month follow-up, a total of 335 breast cancers were confirmed in the final cohort, and the question was: which type of mammography better predicted these cases of cancer? 

We analyzed the data from \citet[][Table 3]{AP-breast-example}.
The AUC and the AP for the two technologies are given in Table~\ref{tab:breast-rslt}, together with various standard error estimates of the AP. 
As in the previous example (Table~\ref{tab:wang-rslt}), the different standard error estimates are in good agreement with each other. 

Overall, digital mammography fared slightly better than film mammography on the AUC-scale, but the AP favored film mammography slightly over digital mammography. 
Thus, depending on the performance measure, we could arrive at different conclusions about which technology was more effective for detecting breast cancer.
The difference in AP (or in AUC) between the two types of mammography was relatively small. We could {\em not} test formally whether these small differences were statistically significant because the malignancy scores based on digital mammograms and those based on film mammograms were not independent, as the mammograms were taken on the same group of patients and rated by the same group of radiologists.
While
\beqnn
\var(\ap_{film} - \ap_{digital}) 
= \var(\ap_{film}) + \var(\ap_{digital}) - 2\cov(\ap_{film}, \ap_{digital}),
\eeqnn
estimating the covariance term, even by the bootstrap, would require us to have information about which pair of scores --- one from digital and another from film mammography --- was for the same patient. We didn't have such information but, given the context, it was safe to conjecture that the covariance term most likely would have been positive.
Hence, by using the standard error estimates in Table~\ref{tab:breast-rslt} and letting $\rho$ be the (unknown) correlation coefficient between $\ap_{film}$ and $\ap_{digital}$,
we can estimate the standard error of the difference, $\ap_{film} - \ap_{digital}$, as a function of $\rho$: 
\beqnn
se(\ap_{film} - \ap_{digital}) 
\approx \sqrt{(0.02)^2 + (0.02)^2 - 2\rho(0.02)(0.02)} 
\approx \begin{cases}
0.020, & \mbox{if} \quad\rho=0.5; \\
0.015, & \mbox{if} \quad\rho=0.7; \\
0.009, & \mbox{if} \quad\rho=0.9.
\end{cases}
\eeqnn
These simple calculations suggest that a difference of about $0.166-0.144=0.022$ on the AP-scale can still be statistically significant if the correlation between digital and film mammography is relatively high, which most likely is the case in reality.

The DMIST publication \citep{AP-breast-example} used the AUC as the main performance measure. Despite their enthusiasm about the effectiveness of digital mammography, the U.S.~Preventive Services Task Force recently concluded that 
``[e]vidence is lacking for benefits of digital mammography and MRI of the breast as substitutes for film mammography'' \citep{breast-rec}. The assessment provided by the AP seems to be in line with the latter conclusion.  

\section{Discussions}
\label{sec:implications}

We have shown in Section~\ref{sec:theory} that, relative to the AUC, the AP places additional emphasis on the {\em initial} true positive rate --- or simply momentum, as we have defined it in Section~\ref{sec:newdef}. In practice, when do we care more about the momentum of a test? We think the momentum is especially important when $\pi$ is relatively small, that is, when the prevalence is low. This is because, when the prevalence is low, we naturally would like to avoid raising too many red flags, but for the precious few flags that we do raise (i.e., the few top-ranked cases), we'd like to have as many true positives as possible. 

Most medical {\em screening} tests do indeed operate under such circumstances, i.e., low prevalence, because the purposes of these tests are to identify diseases in their early stages, when no symptoms are present, so as to facilitate early intervention with the hope to improve outcomes \citep{screening-early-detection}. Therefore, they target a general, asymptomatic population \citep{screening-test}, among which the prevalence is typically very low. By contrast, most {\em diagnostic} tests are aimed at patients who already display some kind of symptoms, so they typically are meant for situations where the prevalence is considerably higher. For assessing screening (as opposed to diagnostic) tests, therefore, a performance metric that emphasizes the test's momentum, such as the AP, may be more attractive than a metric that treats momentum and stamina as being equally important, such as the AUC. In the breast cancer example (Section~\ref{sec:breast}), the disease prevalence was $0.0078$ at 15 months post-screening. In practice, radiologists do consider these very low prevalence numbers when assigning malignancy scores so as to avoid too many false positives \citep{radiology-practice}. In other words, precision is of particular interest for screening, and clinicians may very well prefer a screening test that is favored by the AP to one that is favored by the AUC.

\begin{table}[t]
\centering
\mycap{Prostate cancer example (Section~\ref{sec:prostate}). A simple thought experiment showing changes in the AUC and in the AP as a result of artificially inflating the number of control subjects ($n_0$).}
\label{tab:inflating-controls}
\fbox{%
\begin{tabular}{lr|ccc|ccc} 
& & \multicolumn{3}{c|}{AUC} & \multicolumn{3}{c}{AP} \\
\multicolumn{2}{l|}{Biomarkers}  
& $n_0 \times 1$ & $n_0 \times 10$ & $n_0 \times 100$
& $n_0 \times 1$ & $n_0 \times 10$ & $n_0 \times 100$ \\
\hline
A & 8355.562 & 0.849 & 0.783 & 0.783 & 0.856 & 0.606 & 0.571 \\
  & 7819.751 & 0.850 & 0.857 & 0.857 & 0.802 & 0.370 & 0.062 \\
\hline
B & 8141.232 & 0.810 & 0.773 & 0.773 & 0.850 & 0.572 & 0.468 \\
  & 5074.164 & 0.886 & 0.869 & 0.869 & 0.833 & 0.306 & 0.043 
\end{tabular}}
\end{table}

Sometimes, we may have conducted a case-control study (for which $\pi \approx 50\%$ by design), but would like to use the case-control data to identify biomarkers for the purpose of performing future screening tests (for which $\pi$ is expected to be much smaller). In such applications, it also may be much better to use the AP rather than the AUC to assess the potential biomarkers. In the prostate cancer example (Section~\ref{sec:prostate}), biomarkers were evaluated under a case-control design ($n_1 = 83 \approx 82 = n_0$) for their potential as screening tools for prostate cancer. To see how their relative evaluations would change if the prevalence were much lower, we conducted a simple thought experiment on the two pairs of biomarkers shown in Figure~\ref{fig:prostateROC}, and examined what would happen if we artificially inflated the number of control subjects by creating copies of each existing control subject in the data set (Table~\ref{tab:inflating-controls}). For pair A $(8355.562, 7819.751)$, we already have seen in Section~\ref{sec:prostate} that the two scored similarly on the AUC-scale despite having very different ROC curves --- a clear case of momentum-stamina tradeoff, and that the marker $8355.562$ scored higher on the AP-scale due to its large momentum. Here, we can see that the difference between the two markers becomes even more dramatic on the AP-scale when the prevalence is reduced. For pair B $(8141.232, 5074.164)$, we see that, even though the marker $5074.164$ scored higher on the AUC-scale, based on the case-control data, the AP is more or less indifferent between the two, but, when the prevalence is reduced, the AP can actually start to favor the other marker $8141.232$ by a substantial margin.

Finally, we think the AP is useful not only for medical screening tests, but also for the risk prediction of low probability events in general. Often, models are constructed and covariates are selected in order to predict some future event in a specific population, e.g., the risk of having a cardiovascular event in the next 10 years, or the risk of having a secondary neoplasm in the next 10 years for childhood cancer survivors, and so on. One of the main objectives is to identify patients who have a high risk of developing these conditions. Since many of these events have low probabilities, meaning that the incidence rate is low, the AP may be a better performance measure than the AUC for reasons similar to those discussed above. Currently, however, prediction models and competing risk factors are almost exclusively assessed by ROC curves and more specifically, by the AUC \citep{AUC-exclusive}. 

\section*{Acknowledgments}

\if\blind1
Omitted for blinded version.
\else
WS's research is partially supported by the MacEwan faculty professional development fund.
YY's research is partially supported by the M.~S.~I.~Foundation of of Alberta.
MZ's research is partially supported by the Natural Sciences and Engineering Research Council (NSERC) of Canada.
\fi

\appendix 

\section{Proof of Proposition~\ref{prop:basic}}
\label{sec:prop-proof}

\bitem

\item[(a)] This follows from the very definition of the hit curve. 
Initially ($t=0$), no subject is declared positive (belonging to class-1), so 
$h(0)=0$. In the end ($t=1$), every subject is declared positive including 
all true positives, so $h(1)=\pi$.

\item[(b)] This follows from the fact that, going from $t$ to $t+\Delta 
t$, the worst case is that no additional true positives are identified, and the 
best case is that every subject identified is a true positive.
That is,
\[
 0 \leq h(t+\Delta t) - h(t) \leq \Delta t 
\quad\Rightarrow\quad
 0 \leq \frac{h(t+\Delta t) - h(t)}{\Delta t} \leq 1
\]
for all $\Delta t > 0$. Taking the limit on both 
sides gives
\[
 0 \leq \lim_{\Delta t \rightarrow 0} 
 \frac{h(t+\Delta t) - h(t)}{\Delta t} \leq 1,
\]
or $0 \leq h'(t) \leq 1$. 

\item[(c)]
Integration by parts implies
\[
 \int h(t) dh(t) = 
 \left[ h^2(t) \right]_{0}^1
-\int h(t) dh(t).
\]
Solving for $\int h(t) dh(t)$ gives
\[
 \int h(t) dh(t) = \frac{h^2(1)-h^2(0)}{2} = \frac{\pi^2}{2},
\]
by Proposition~\ref{prop:basic}(a).
\eitem

\section{Asymptotic variance of the AP: Some details}
\label{sec:APvar-details}

Due to the multinomial constraints, $p_1+...+p_K=1$ and $q_1+...+q_K=1$, in practice we work with $(K-1)$-dimensional vectors, $\myvec{p}$ and $\myvec{q}$,
rather than $K$-dimensional vectors.
By direct algebraic calculations, we can obtain that
\[
\widehat{\myvec{J}} =
\left[
\begin{array}{ccc}
\myvec{P} & - & - \\
- & \myvec{Q} & - \\
- & - & a
\end{array}
\right],
\]
where $\myvec{P}$, $\myvec{Q}$ are $(K-1)\times(K-1)$ matrices with
\[
\myvec{P}_{kk}  = \frac{z_k}{\wh{p}_k^2} + \frac{z_K}{\wh{p}_K^2} \quad\mbox{for}\quad k=1,...,K-1,
\quad
\myvec{P}_{kk'} = \frac{z_K}{\wh{p}_K^2} \quad\mbox{for all}\quad  k \neq k',
\]
\[
\myvec{Q}_{kk}  = \frac{\bar{z}_k}{\wh{q}_k^2} + \frac{\bar{z}_K}{\wh{q}_K^2} \quad\mbox{for}\quad k=1,...,K-1,
\quad
\myvec{Q}_{kk'} = \frac{\bar{z}_K}{\wh{q}_K^2} \quad\mbox{for all}\quad  k \neq k',
\]
and
\[
 a = \frac{n_1}{\wh{\pi}^2}+\frac{n_0}{(1-\wh{\pi})^2}.
\]
Now, let
\[
P_k \equiv \sum_{k' \leq k} \wh{p}_{k'},\quad
Q_k \equiv \sum_{k' \leq k} \wh{q}_{k'},\quad\mbox{and}\quad
C_k \equiv \wh{\pi}P_k+(1-\wh{\pi})Q_k.
\]
Again, by direct algebraic calculations, we can obtain that 
\[
\nabla g = 
\left[
\begin{array}{c}
\nabla_p \\
\nabla_q \\
\nabla_{\pi} 
\end{array}
\right]
\]
where $\nabla_p$, $\nabla_q$ are $(K-1)$-dimensional vectors with
\[
\nabla_p(k) \equiv \frac{\partial g}{\partial \wh{p}_k} = 
\frac{\wh{\pi}P_k}{C_k} + 
\sum_{k'=k}^{K-1}{ 
\wh{p}_{k'} \left[ \frac{\wh{\pi}(1-\wh{\pi})Q_{k'}}{C_{k'}^2} \right]} - \wh{\pi}, 
\]
\[
\nabla_q(k) \equiv \frac{\partial g}{\partial \wh{q}_k} = 
\sum_{k'=k}^{K-1}{ 
\wh{p}_{k'} \left[ \frac{-\wh{\pi}(1-\wh{\pi})P_{k'}}{C_{k'}^2} \right]},
\]
for $k=1,..., K-1$, and
\[
\nabla_{\pi} \equiv \frac{\partial g}{\partial \wh{\pi}} =
\sum_{k=1}^K
\wh{p}_k \left[ \frac{P_k Q_k}{C_k^2} \right].
\]

\bibliographystyle{/u/m3zhu/natbib}
\bibliography{./ref}

\end{document}